\documentclass[12pt, preprint]{emulateapj}
 \slugcomment{Accepted to ApJL}
\usepackage{apjfonts}
 \newcommand{\Jnic}{$J_{110}$}
 \newcommand{\Hnic}{$H_{160}$}
 
 \newcommand{\lya}{Ly$\alpha$}
 
 \newcommand{\jdtwo}{JD2325+1433}

\begin{document}

\title{A Lyman Break Galaxy Candidate at $\lowercase{z} \sim9$\altaffilmark{1}}
 
 \author{Alaina L. Henry\altaffilmark{2},  Matthew A. Malkan\altaffilmark{2}, James W. Colbert\altaffilmark{3}, 
Brian Siana\altaffilmark{3}, Harry I. Teplitz\altaffilmark{3},  Patrick McCarthy\altaffilmark{4}}
 
 \altaffiltext{1}{
 This work is based in part on observations made with the Spitzer Space Telescope, which is operated by the Jet Propulsion Laboratory, California Institute of Technology under a contract with NASA. Support for this work was provided by NASA through an award issued by JPL/Caltech.
  This work is also based in part on observations made with the NASA/ESA {\it Hubble Space Telescope}, obtained from the Space Telescope Science Institute, which is operated by the Association of Universities for Research in Astronomy Inc., under NASA contract NAS 5-26555.  These observations are associated with proposals 9484, 9865, 10226, and 10899.}

 \altaffiltext{2}{Department of Physics and Astronomy, Box 951547, UCLA, Los Angeles, CA 90095, USA; ahenry@astro.ucla.edu                   }
 \altaffiltext{3}{$Spitzer$ Science Center, California Institute of Technology, 220-6, Pasadena, CA, 91125, USA}
 \altaffiltext{4}{Observatories of the Carnegie Institute of Washington, Santa Barbara Street, Pasadena, CA 91101}

 \begin{abstract}
We report the discovery of a $z\sim9$ Lyman Break Galaxy (LBG) candidate,  selected from 
the NICMOS Parallel Imaging Survey as a J-dropout with \Jnic\ - \Hnic\ = 1.7. {\it Spitzer}/IRAC photometry reveals that the galaxy has a blue \Hnic\ - 3.6 \micron\ color, and a spectral break between  3.6 and 4.5 \micron.   We interpret this break as the Balmer break, and  derive a best-fit photometric redshift of $z\sim9$.  
  We use Monte Carlo simulations to test the significance of this photometric redshift, and show a 96\% probability of $z\ge7$.   We estimate a lower limit to the  comoving number density of such galaxies at $z\sim9$ of $\phi > 3.8 \times 10^{-6} ~ {\rm Mpc}^{-3}$.    If the high redshift of this galaxy is confirmed, this will indicate that the luminous end of the rest-frame UV luminosity function has not evolved substantially from $z\sim 9$ to $z\sim3$. 
Still, some small degeneracy remains between this $z\sim9$ model and models at $z\sim2-3$; deep optical imaging (reaching $I_{AB} \sim 29$) can rule out the lower-z models.  
 \end{abstract}

 \section{Introduction}

The search for the first galaxies remains one of the primary goals of extragalactic astronomy. 
Surveys of distant star forming galaxies (e.g. Lyman break galaxies; LBGs) 
have resulted in determinations of luminosity functions, stellar masses, and ages of 
objects to  $z \sim 6$ (\citealt{Stanway03}; \citealt{bunker}; \citealt{Yan06}; \citealt{Yoshida}; \citealt{Bouwens06}; \citealt{Eyles}; \citealt{Stark07}). 
The space density of  
galaxies at $z\sim6$ and their apparent maturity suggests that galaxy formation must
have begun at $z> 7$. 
 Although the requirement of deep near-infrared imaging makes $z\ge7$ difficult to observe,  this is a critical epoch 
in the history of galaxy formation.
Discovery of the first galaxy-wide bursts of star formation--the first genuine protogalaxies--is essential to understanding  galaxy evolution.

Pencil beam surveys have searched for $z\ge8$ galaxies.  In the deepest available near-infrared data---the Hubble Deep Fields North and South (HDF-N and -S), the Ultra Deep Field (UDF) and the UDF Parallel Fields---\cite{Bouwens05} find three possible $z\sim10$ galaxies.  In addition,  observations of gravitationally lensed galaxies have also revealed a few potential $z\ge8$ galaxies (Richard et al.~2006, 2008).  However, these deep,  small area surveys find faint candidates which are difficult to follow up, so none of the redshifts have been confirmed by measuring their spectral energy distributions (SEDs) at longer wavelengths. This crucial test has ruled out other $z\sim10$ candidates  \citep{Bouwens05}, including the one first presented in \cite{Dickinson00}.

Our NICMOS Parallel Imaging Survey has the unique combination of sensitivity and area to find rare, luminous J-dropout LBGs at $8\le z \le 10$.   These candidates are typically  \Hnic\ = 23 - 25 (AB), so follow-up observations can confirm or rule out extremely high redshifts.   In particular, these galaxies can be observed with IRAC on {\it Spitzer Space Telescope} (hereafter {\it Spitzer}).   We present here the first results of such a survey, where one possible $z\sim 9$ galaxy has been discovered.  We use $H_0 = 70 ~{\rm   km s^{-1}~ Mpc^{-1}}$, $\Omega_{\Lambda} = 0.7$, and $\Omega_M = 0.3$, and AB magnitudes are also used throughout.

 \section{Observations \&  Data Reduction}
 \label{targets}
 We selected J-dropout candidates from the NICMOS Parallel Imaging Survey ( $\sim 135$ arcmin$^{2}$; \citealt{Teplitz}; \citealt{Yan00}; \citealt{Colbert_xray}; \citealt{Henry07}).   
 In this paper, we have adjusted the photometry by -0.16 and -0.04 mag in \Jnic\ and \Hnic\, according to de Jong (2006).  From more than seven thousand detected sources, four targets were selected from the small subset (2\%) with red \Jnic\ - \Hnic\ ($>0.9$) and no \Jnic\ detection above $2 \sigma$.  
 
 \begin{figure*}
\epsscale{0.9} 
\begin{center}
\plotone{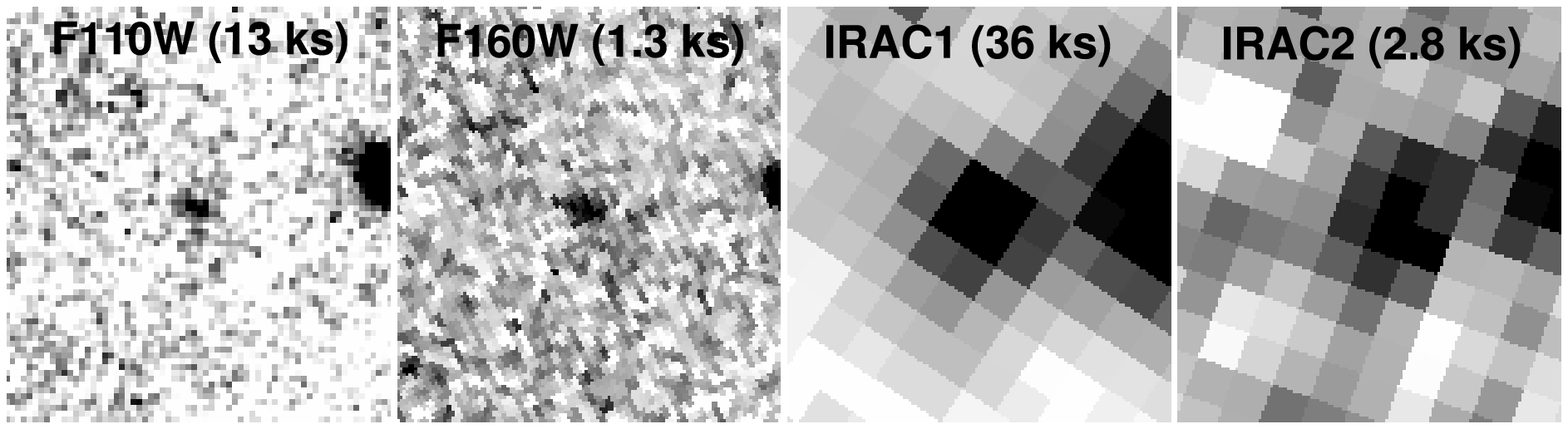}
\end{center}
\caption{ {\bf Postage stamp images of \jdtwo\ show that it is elongated in the NICMOS images. } Images are 6\arcsec\ on a side.  Note that the integration in  \Jnic\ is 10 times longer  than  in \Hnic, and the integration in IRAC1 (3.6 \micron) is 13 times longer than in IRAC2 (4.5 \micron), so the galaxy is indeed red in \Jnic\  - \Hnic, and [3.6]-[4.5].  The coordinates of \jdtwo\ are 23h24m46.52s, 14d32m594s, bootstrapped to the 2MASS frame. }
\label{stamps}
\end{figure*}

We observed these NICMOS selected J-dropout candidates for four, five or six orbits  each with NICMOS  in the \Jnic\ -band to confirm their red \Jnic\ - \Hnic\ color.   The  images were drizzled according to \cite{drizzle}, and the resulting point-spread function has a FWHM$\sim 0.3\arcsec$.   The 5 $\sigma$ sensitivity, measured in a 0.8\arcsec\ diameter aperture is  \Jnic\ =  26.3- 26.5.    
The NICMOS colors were measured in apertures of this size.

These four galaxies  were also observed with the Infrared Array Camera (IRAC) on {\it Spitzer}, at 3.6, 4.5, 5.8, and 8.0 \micron.  The integration times were 0.9 ks pixel$^{-1}$ at 3.6  and 5.8\micron, and 2.8 ks pixel$^{-1}$  at 4.5 and 8.0 \micron.  We used the {\it Spitzer} Science Center mosaicking software, MOPEX, with the drizzle algorithm to combine the images from the Basic Calibrated Data set.  The output pixel size is 0.6\arcsec, and resulting FWHM of the point-spread-function is 1.8\arcsec\ in the 3.6 and 4.5 \micron\ bands, 2.0\arcsec\ in the 5.8 \micron\ band, and 2.2\arcsec\ in the 8.0 \micron\ band.  The noise was measured by placing apertures in blank parts of the images.   We chose a 2.4\arcsec\ diameter aperture to avoid contamination from nearby sources.   For this small aperture, the corrections are 1.8, 1.9, 2.4, and 2.7 in the four IRAC bands (derived from point sources in our fields).  The 
resulting  3 $\sigma$ sensitivities are 0.8, 0.8, 5.8, and 3.8 $\mu$Jy (total).  

\section{Results}
The resulting photometry shows that three of the four sources have SEDs which are consistent with lower-z interlopers.  One has \Jnic\ - \Hnic\ = 0.4, and it too blue to be a viable $z>8$ candidate.  Two other sources have rising SEDs into the IRAC bands, with \Hnic\ - [3.6]  = 2-3.  This makes them likely lower redshift interlopers, such as dusty starburst galaxies at $z\sim2-4$ (e.g. \citealt{cm04}; \citealt{Labbe05}; \citealt{brammer}; \citealt{Stutz}).   The most remarkable source, however, is \jdtwo\ which has \Jnic\ -\Hnic\ = 1.7, and a spectrum which is flat (in $f_{\nu}$) through 3.6 and 4.5 \micron\ (much bluer than the controversial $z\sim 6.5$ J-dropout candidate reported by \citealt{Mobasher}).  These are the colors that are expected for LBGs at $8 < z < 10$. 

Initially,  \jdtwo\ was only detected at 4.5 \micron\ ($f_{\nu} = 1.29 \pm 0.29~\mu$Jy), even though the 3.6 and 4.5 \micron\ images reach the same sensitivity.    Since this could be evidence for a Balmer break  at a redshift of $z\sim9$, we obtained deep (10 hour)  {\it Spitzer} Directors Discretionary Time (DDT) observations at 3.6  and 5.8 \micron\  to verify the break.  The data are reduced in the same manner as described above, and the 3 $\sigma$ sensitivity is 0.23 and 2.0 $\mu$Jy  at 3.6 and 5.8 \micron. The candidate, \jdtwo\ is detected at 3.6 \micron, with $f_{\nu}  = 0.75 \pm 0.08~ \mu$Jy, confirming this break between 3.6 \micron\ and 4.5 \micron.  We do not detect \jdtwo\ at 5.8 \micron, which is consistent with its flat spectrum.  
 
\begin{figure*}
\plottwo{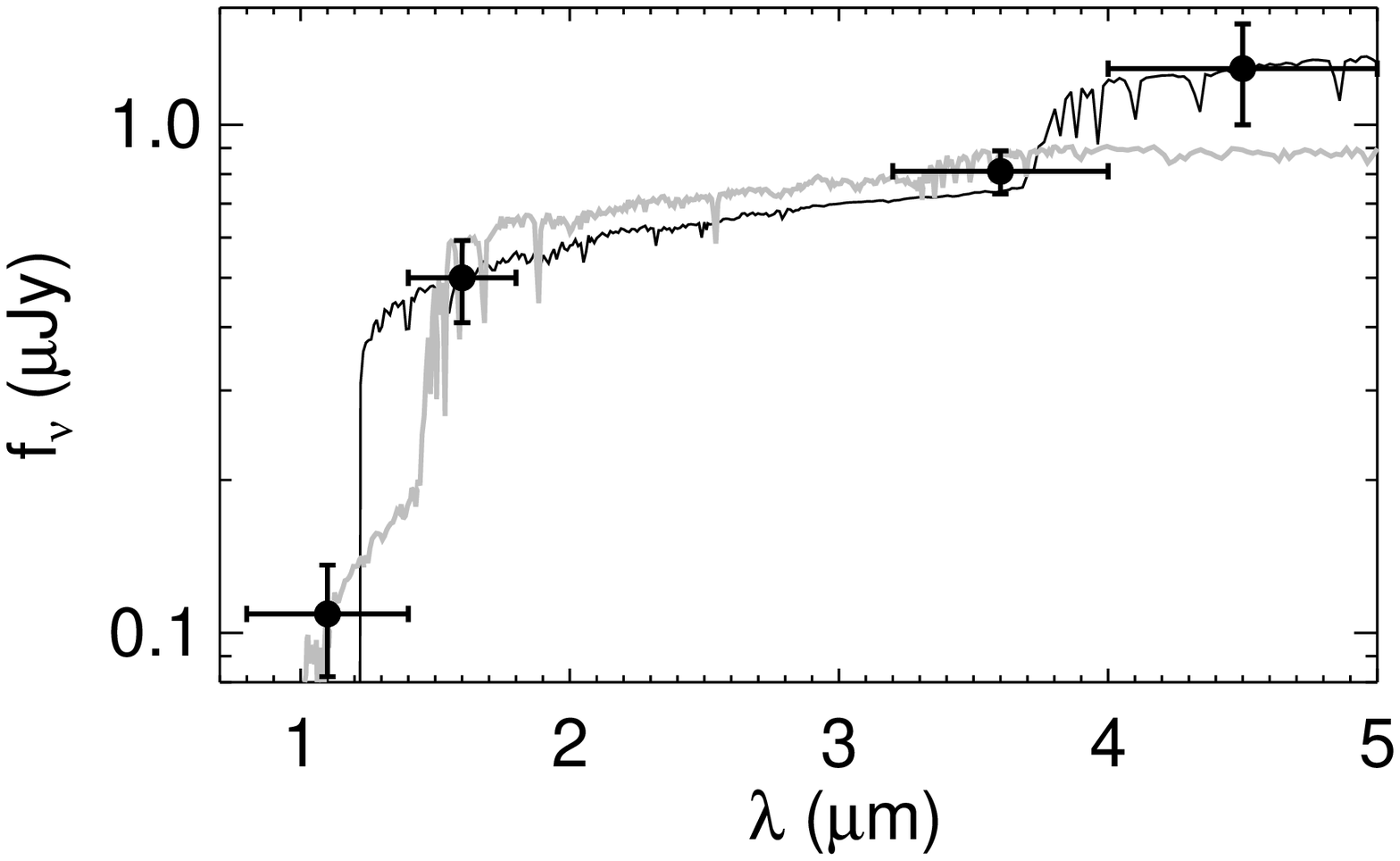}{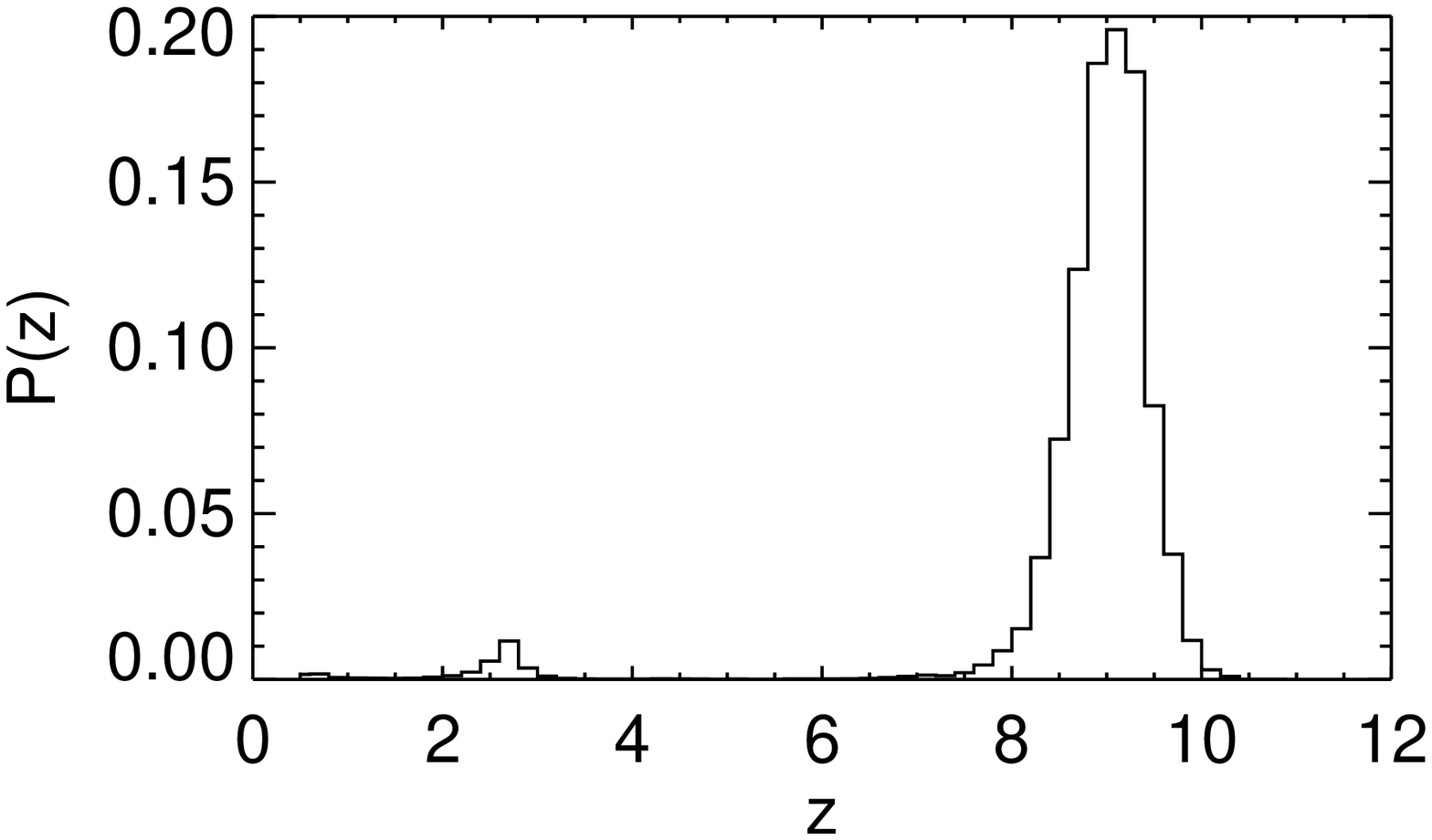}
\caption{ {\it Left--} {\bf The photometry of \jdtwo\ shows two breaks in the spectrum,}  consistent with the Lyman and Balmer breaks and suggesting a redshift of $z\sim9$.   The black curve is the best fit model ($z=9.0$ with an age of 64 Myrs, a constant SFR of 930 $M_{\sun}~{\rm yr}^{-1}$,  $A_V = 1.0$, $M_B = -23.8$, $M=5.4\times 10^{10}~ {\rm M_{\sun}}$ and $Z = 0.2Z_{\sun}$).  The grey curve is an alternate, poor fit at $z=2.9$ (an instantaneous burst aged 360 Myrs, with $A_V = 0.1$, $M_B  = -21.1$, $M = 1.8 \times 10^{10}~ {\rm M_{\sun}}$ and $Z = Z_{\sun}$). 
  {\it Right--} The results of our Monte Carlo simulation show probability as a function of redshift.   More than 96\% of solutions lie at $z\ge7$.   }
  \label{fits}
\end{figure*}

Postage stamp images of \jdtwo\ in Figure \ref{stamps} show another galaxy approximately 2.5\arcsec to the east.   To test whether there is significant contamination from this source, we performed a PSF subtraction to remove it.  This assumes that the nearby source is unresolved the IRAC images, which is consistent with its NICMOS size  (FWHM$\sim 0.5$\arcsec) and the observed IRAC FWHM ($\sim 1.8$\arcsec).  We then re-measure \jdtwo\ with the nearby source removed.  The resulting flux density is $f_{\nu} = 0.82~ {\rm and}~ 1.25~ \mu$Jy at 3.6 and  4.5 \micron.  
This is consistent with the expectation that at most 2-3\% of the light from a point source at the position of the nearby source would fall in the aperture that we used for \jdtwo--- a fraction that corresponds to at most 0.05 $\mu$Jy.  

As an independent check, we used SExtractor  \citep{sextractor} to deblend the two sources.  The resulting photometry, using the same 2.4 \arcsec\ diameter aperture, is consistent with aperture photometry when no correction for blending is made---$f_{\nu} = 0.81~ {\rm and}~ 1.33 ~\mu$Jy at 3.6 and 4.5 \micron.     Therefore, we conclude that contamination is negligible, and make no correction to the aperture photometry.   We do, however, include the scatter derived from these three measurements in our IRAC error estimates, although it is only appreciable for the 3.6 \micron\ photometry (0.04 $\mu$Jy, added in quadrature with 0.08 $\mu$Jy).

\subsection{Matching IRAC \& NICMOS Photometry} 
\label{matching}
 To facilitate comparison with IRAC, we must estimate an aperture correction to account for missed flux in the NICMOS images.  
Although \jdtwo\ does not have the S/N required to characterize its morphology, we can place rough constraints on its size. 
We use GALFIT \citep{Peng}  to derive sizes and axial ratios for a few different assumed light profiles.    For an exponential disk model, we find $r_s = 0.2 \pm 0.03$ \arcsec\ and $a/b = 0.36 \pm 0.09$, and for a Gaussian we find a  FWHM $= 0.53 \pm 0.06$ \arcsec, and $a/b =0.36 \pm 0.11$.      We also attempted to fit a de Vaucouleurs profile, but it was unconstrained.    
The uncertainties on these fits define the largest and smallest galaxy sizes, and therefore the largest and smallest aperture corrections that we can expect. 
To account for a possible de Vaucouleurs light profile, we add 0.15 magnitudes of missed flux to the largest exponential disk aperture correction as found by \cite{Colbert}.   
We choose an aperture correction in the center of this range, and 1$\sigma$ uncertainties to span the entire range.     For a 0.9\arcsec\ diameter aperture, this aperture correction is $1.18 \pm 0.12$.   We find that this aperture size minimizes the uncertainty on the total NICMOS flux density, which includes uncertainties for both the background and the aperture correction.   We apply this aperture correction to \Jnic\ and \Hnic, since the PSF is approximately the same for both.  

Finally, the {\it total}  flux densities are:  (\Jnic, \Hnic,  [3.6], [4.5]) = (0.11, 0.50, 0.75, 1.29) $\pm$ ( 0.03, 0.09, 0.09, 0.29) $\mu$Jy.    The 3$\sigma$ upper limits at 5.8 and 8.0 \micron\ are  2.0 and 3.8 $\mu$Jy.

\subsection{On the Possibility of a Stellar Contaminant}
A stellar or substellar contaminant for \jdtwo\ is improbable, for a several reasons.  First, the source is likely resolved.  In \S \ref{matching} we show that the semi-major axis has a FWHM $\sim$ 0.5\arcsec, while the NICMOS PSF has a FWHM of  $\sim 0.35$\arcsec.  Furthermore, the NICMOS images in Figure \ref{stamps} show that \jdtwo\ is elongated in the same direction in both the \Jnic\ and \Hnic\  images.

Even if \jdtwo\ were unresolved, its photometry can not be fit by a stellar contaminant.  Of particular concern are brown dwarfs, which have molecular features that can imitate spectral breaks.  While the reddest L-dwarfs can fit our NICMOS color, they can not simultaneously fit our IRAC photometry (typically [3.6] - [4.5] $\la -0.5$).  Similarly, T-dwarfs can fit our IRAC photometry with  [3.6] - [4.5] $> 0.5$ (AB),  but with   J - H $< 0$, they are much too blue to fit our NICMOS observations (\citealt{Burgasser}; \citealt{Patten}).   This trend in the colors of  late-type T-dwarfs is predicted to extend to the coolest dwarfs (\citealt{baraffe}; Barman, priv. comm.).  Lastly, we note that other young stellar objects such as T-Tauri stars can not produce this red \Jnic\ - \Hnic\ color (e.g. \citealt{Hernandez}), as their NIR colors are dominated by photospheric emission.

\section{Constraining the Redshift of \jdtwo}
\label{hyperz}
We use the {\it Hyperz} photometric redshift code \citep{Bolzonella} with \cite{bc03} stellar snythesis templates to estimate the redshift of  \jdtwo.  We do not include the 5.8 \micron\ limit, as no reasonable models predict a flux $\ga 2~ \mu$Jy at 5.8 \micron.  Assuming a Salpeter initial mass function (IMF) and using the Padova 1994 evolutionary tracks, we fit for redshift, varying age, mass normalization, extinction ($A_V = 0-3$; \citealt{Calzetti} law),  and metallicity ($Z = 0.005, 0.02, 0.2, 0.4, 1.0 \times Z_{\sun}$).  We also allow three different star formation histories--  an instantaneous  burst, a constant star formation rate (SFR), and an exponentially declining SFR with an e-folding time of 100 Myrs.  With four data points, the problem is underconstrained, and we are not able to estimate metallicity or star formation history.  However, we include these parameters so that we do not overlook some combination of parameters that is a good fit to our data at lower redshift ($z\sim2-4$).

In Figure \ref{fits} we compare the best fit model (black curve), which has $z=9.0$, to the best possible fit that can be achieved for $z\le6$ (grey).  This lower-z fit, which has $z= 2.9$, is significantly poorer, and can not reproduce the two breaks in the spectrum.  To further test the significance of these fits, we generated $10^5$ Monte Carlo realizations of the photometry, simultaneously varying the fluxes according to the measured uncertainties.  We fit these data with {\it Hyperz}, as described above.  The probability distribution in redshift space (see Figure \ref{fits}, right), shows a peak at $z=9.0$ and a much smaller peak at $z\sim2.5$.  More than 96\% of solutions lie at $z\ge7$.  This is consistent with the observed 4.5 \micron\ flux density lying  only 1.7$\sigma$ away from the $z=2.9$ model.  For normally distributed uncertainties, the true 4.5 \micron\ flux density  will lie more than 1.7$\sigma$ away from our measurement 10\% of the time.   This implies that there is a 5\% chance that the Balmer break is more than 1.7 $\sigma$ larger than we have estimated, and 5\% chance that there is no Balmer break at all.  In the latter case, $z\sim2-3$ models will be a good fit to the photometry.

\begin{figure*}
 \plottwo{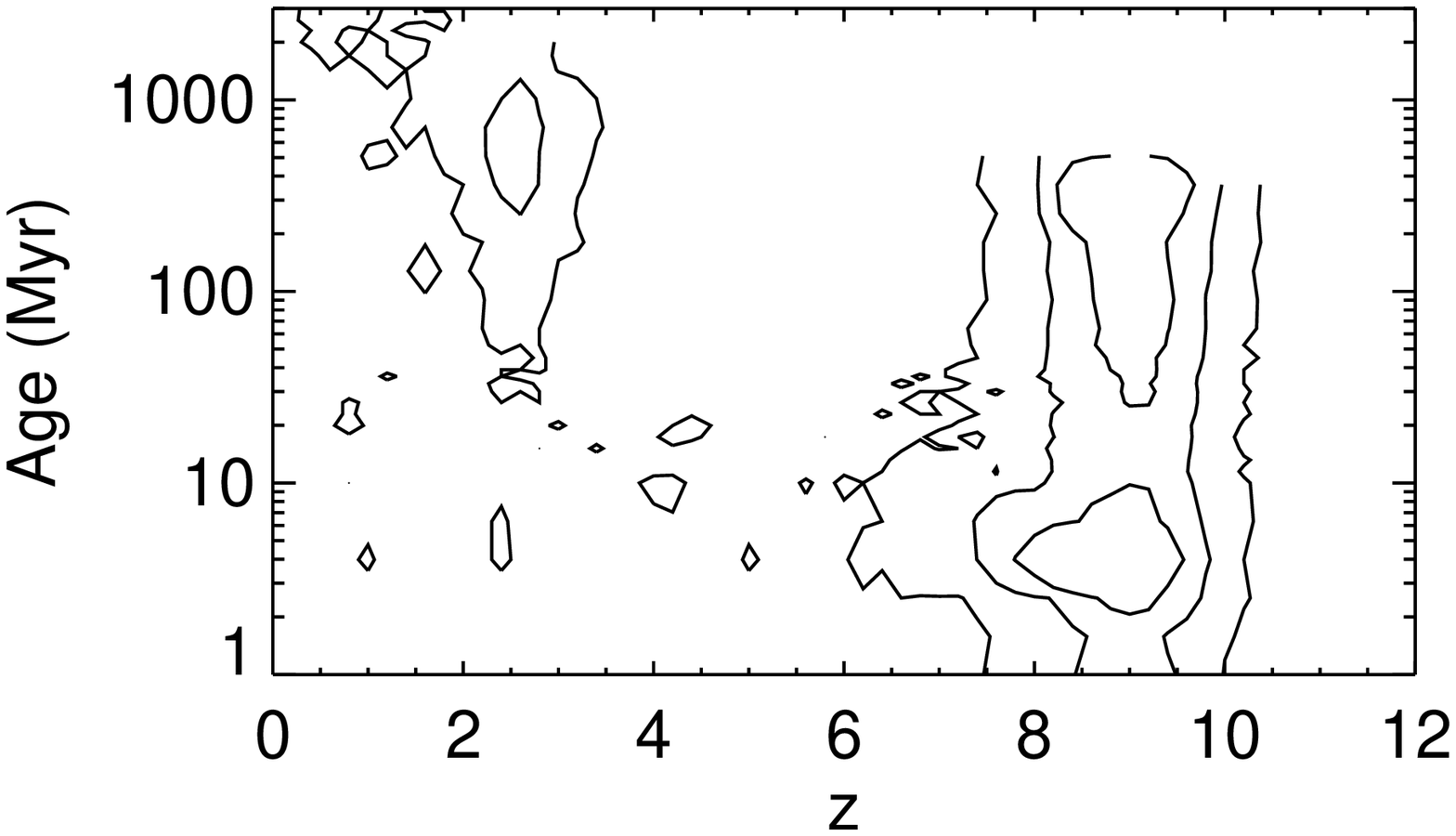}{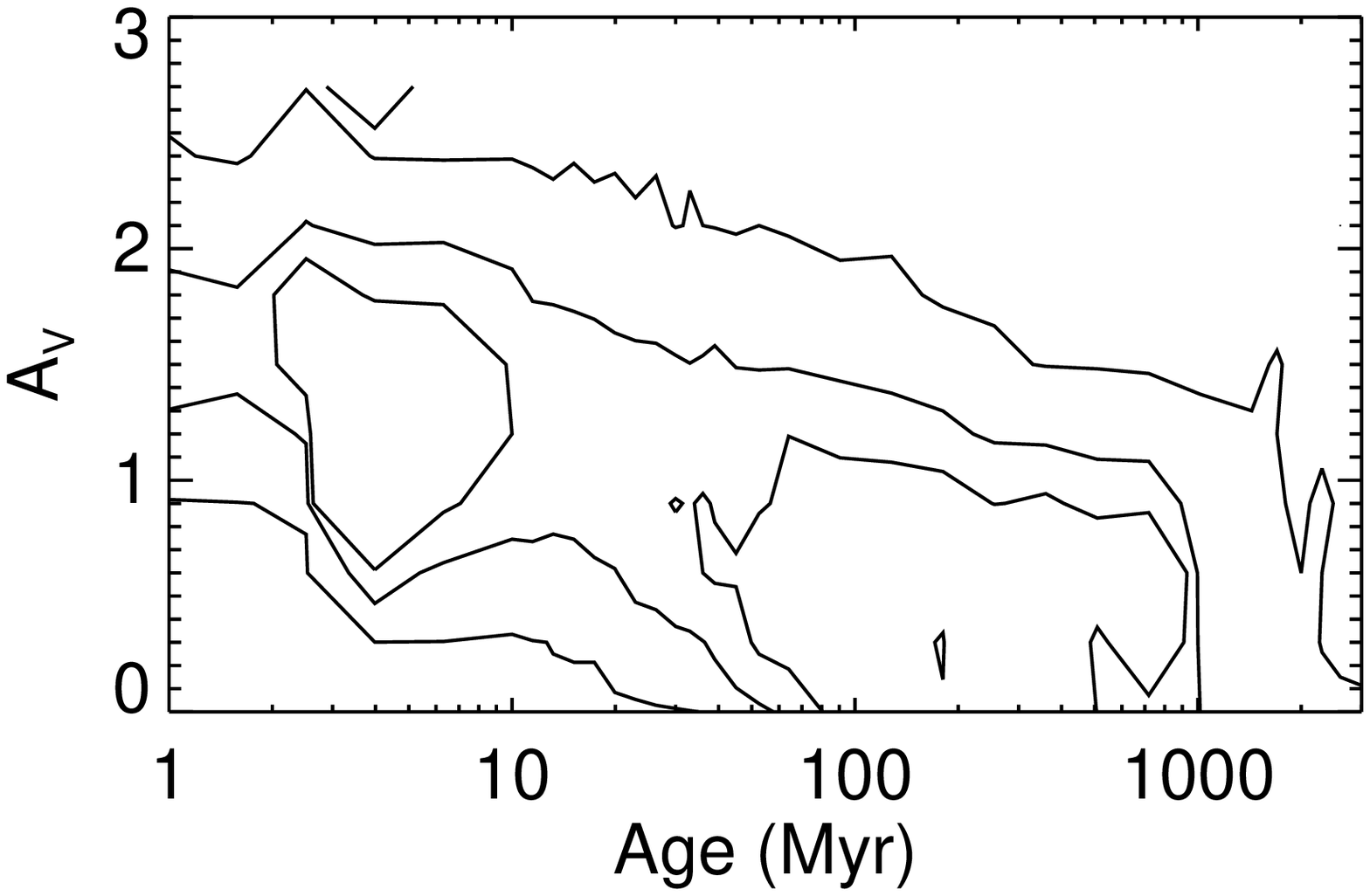}
\caption{Confidence intervals show that a $z\sim9$ solution is preferred at the 1$\sigma$ level, while age and extinction are not strongly constrained.   Contours are 1, 2, and 3 $\sigma$.   On the left hand panel, contours are open because best-fit ages are restricted to be younger than the age of the universe. }
\label{contours}
\end{figure*}

In Figure \ref{contours} we show two dimensional confidence intervals in age/redshift space, and extinction/age.   Estimated 1, 2, and 3 $\sigma$ contours are determined from the  probability at which 68.3\% (95.4\%, 99.7\%) of the Monte Carlo realizations have  a higher likelihood.  In the left panel we show that the $z\sim9$ solution is preferred at the 1 $\sigma$  level, while the 2 $\sigma$ interval includes some $z\sim2.5$ solutions.  Both panels show that the age of \jdtwo\ is not well constrained, with values ranging from a few Myrs,  to several hundred Myrs.   The right panel shows a degeneracy between age and extinction.  
If we exclude models with ages less than 10 Myrs (since it is unlikely that all the stars in the galaxy can form within 10 Myrs), then we can also constrain the extinction to $A_V \la 1$.  Lastly, we note that estimated stellar masses range from $10^{10}$ to $10^{11} ~ {\rm M_{\sun}}$,  for a Salpeter IMF.

\section{Implications}

\label{implications}
If \jdtwo\ is at $z\sim9$, its UV luminosity corresponds to $5L^*(z=3)$, and we can estimate a lower limit on the volume density of luminous $8 < z <10$ LBGs.\footnotemark[1]\footnotetext[1]{Since we have not observed all NICMOS Parallel selected J-dropout galaxies with IRAC, several more galaxies like \jdtwo\ may exist in our survey. } For this redshift range,  the entire NICMOS Parallel Survey covers $5.2\times 10^5~ {\rm Mpc}^3$  (comoving) to \Hnic\ = 24.   Beyond this, we are limited by \Jnic\ - band sensitivity required to select sources with \Jnic\ - \Hnic\ $\ge 0.9$, as described in \S \ref{targets}.  At fainter magnitudes (given the large inhomogeneity in pure parallel exposure times), 51\% of the total survey area has sufficient sensitivity to find galaxies like \jdtwo.     This implies that we have searched $2.6\times 10^5~ {\rm Mpc}^3$ (comoving), and the resulting volume density of $8<z<10$ LBGs  is $\phi > 3.8 \times 10^{-6} ~ {\rm Mpc}^{-3}$.
 
 The volume density implied here is  roughly consistent with that at $z\sim3$, as measured by \cite{Steidel99}.  This  LF  predicts $\phi \sim 10^{-6} ~ {\rm Mpc}^{-3}$ for galaxies with $L > 5 L^*(z=3)$, implying $0 - 0.75$ such sources in our survey volume (assuming Poisson errors).   On the other hand,  $z\sim6$ LF, as measured by Bouwens et al. (2006, 2007), predicts fewer luminous galaxies and therefore significantly smaller number for this survey ($< 0.001$).    If this $z\sim6$ LF is representative of $z\sim9$ LBGs, then \jdtwo\ is almost certainly an interloper at $z\sim2-3$.     However, this evolution in the LF is uncertain, and the $z\sim5$ LF measured by \cite{Iwata} shows no change in the number density of luminous galaxies from $z\sim3$.  Since the Iwata et al. survey contains four times more area than the Bouwens et al. survey, this measurement may be a more reliable probe of the most luminous sources. 
 
Indirect evidence also implies that $z\ge8$ galaxies may be detectable in the NICMOS pure parallel imaging.  At $z\sim5-6$, the ages and masses of LBGs predict a more intense period of star formation at earlier times (\citealt{Yan06}; \citealt{Eyles}; \citealt{Stark07}) suggesting upward luminosity evolution at $z>6$.  This argument  is also made for faint \lya\ emitting galaxies.   If any of the six candidates reported by \cite{Stark07_lya} are indeed at $z>8$, then  the space density of faint \lya\ emitters at this epoch must be larger than at $z = 5.7$ (as measured by \citealt{MR04}; \citealt{Shimasaku06}).  
In conclusion, the density of  galaxies and the evolution of the LF at $z>5$ is sufficiently uncertain that the $z\sim9$ interpretation of \jdtwo\ can not be ruled out.

 Constraining the numbers of galaxies at $z>8$ is crucial to understanding the transition from a neutral to ionized intergalactic medium (IGM).  Currently, it is unclear if  star forming galaxies at $z\sim6$ are capable of reionizing the IGM,  largely in part due to uncertainties in the stellar IMF (e.g. \citealt{Chary_imf}), the escape fraction of ionizing photons (\citealt{malkan}; 
 \citealt{Shapley06}; \citealt{Siana}) and the gas clumping factor.   If \jdtwo\ is indeed at $z\sim9$, the relatively large number of UV-luminous galaxies at this redshift will likely relax the requirement for strong evolution in these parameters.

Finally, we note that further observations can resolve the degeneracy between the $z\sim9$ and $z\sim2-3$ interpretations.  
The blue \Hnic\  - IRAC colors constrain the SED so that we can predict  $I_{AB} \sim 27-28$ for the intermediate-z solutions.  Of the 4\% of Monte Carlo realizations (discussed in \S \ref{hyperz}) that have $z < 7$, none are fainter than $I_{AB} = 29$, so deep optical imaging can definitively confirm the redshift  of \jdtwo.

\acknowledgments
 This research was supported by NASA through {\it Hubble Space Telescope} Guest Observer grant 10899.   We thank the {\it Spitzer} director, T. Soifer, for the award of the DDT.  Ned Wright's Javascript Cosmology Calculator was used in preparation of this paper.  We are grateful to C. Papovich for for helpful comments that improved this manuscript.




\end{document}